\documentclass{nature}
\usepackage{graphicx,subfigure,amsmath,bbm}
\usepackage[hyphens]{url}
\usepackage[hidelinks]{hyperref}
\hypersetup{breaklinks=true}

\title{Dichotomy of the Photo Induced 2-Dimensional Electron Gas on SrTiO$_3$ Surface Terminations}

\author{S. N. Rebec$^{1,2,*}$, T. Jia $^{2,3,*}$, H. M. Sohail$^{2}$, M. Hashimoto$^{4}$, D.-H. Lu$^{4}$, Z.-X. Shen$^{1,2,3}$, \& R. G. Moore$^2$}

\begin{document}

\maketitle

\begin{affiliations}
 \item Department of Applied Physics, Stanford University, Stanford, California 94305, USA
\item Stanford Institute for Materials and Energy Sciences, SLAC National Accelerator Laboratory, Menlo Park, California 94025, USA
\item Department of Physics, Stanford University, Stanford, California 94305, USA
\item Stanford Synchrotron Radiation Lightsource, SLAC National Accelerator Laboratory, Menlo Park, California 94025, USA
\item[] $^*$ These two authors contributed equally to the work.
\end{affiliations}

\begin{abstract}
Oxide materials are important candidates for the next generation of electronics due to a wide array of desired properties which they can exhibit alone or when combined with other materials. While SrTiO$_3$  (STO) is often considered a prototypical oxide, it too hosts a wide array of unusual properties including a two dimensional electron gas (2DEG) which can form at the surface when exposed to UV light. Using layer-by-layer growth of high quality STO films, we show that the 2DEG only forms with the SrO termination and not with the TiO$_2$ termination, contrary to expectation. This behavior is similarly seen in BaTiO$_3$ (BTO), in which the 2DEG is only observed for BaO terminated films. These results will allow for a deeper understanding, and better control, of the electronic structure of titanate films, substrates and heterostructures.
\end{abstract}

There are few material systems which exhibit the wide range of relevant physical phenomena as SrTiO$_3$. It is well known for having a high dielectric constant \cite{MullerK.Burkard1979} and undergoing a superconducting transition at 0.3 K\cite{Temperatures1967}. When combined with other materials, either as a layered compound or as a thin film substrate, a much wider array of phenomena are observed. For example, when FeSe, an iron-based superconductor, is grown on STO its superconducting $T_c$ increases from $8\ \text{K}$ to $\geq 60\ \text{K}$ \cite{Wang2012, Lee2014}. In addition, at the interface of insulating SrTiO$_3$ and LaAlO$_3$ a 2DEG emerges, along with superconductivity and ferromagnetism \cite{Ohtomo2004, Caviglia2008 ,Bert2011, Li2011}. A 2DEG can also be generated at the surface of STO alone by exposure to UV light, like that from a synchrotron lightsource \cite{Meevasana2010,Santander-Syro2011, Hatch2015, Walker2015}. The UV light is believed to create oxygen vacancies through a double Auger process \cite{Walker2015}. The remaining electrons create a Ti$^{3+}$/Ti$^{4+}$ mixed valence state and populate the Ti t$_{2g}$ bands\cite{Walker2015}. Existence of subands due to quantum confinement confirms that the excess electrons are trapped at the surface and contribute to the itinerant carrier density and the 2DEG \cite{Santander-Syro2011}. The quick refilling of the oxygen vacancies with low doses of oxygen suggests that the oxygen vacancies are also localized at the surface\cite{Walker2015, Sitaputra2015}.

The 2DEG on the surface of STO has been heavily studied using angle resolved photoemission spectroscopy (ARPES)\cite{Meevasana2010,Santander-Syro2011, Hatch2015, Walker2015, Wang2014, McKeownWalker2016 }. Measurements have been carried out on the (001), (110) \cite{Zhang2017} and (111)\cite{McKeownWalker2014} faces of STO using either fractured single crystal samples or commercially available substrate wafers. The fractured single crystal samples likely have a mixed SrO/TiO$_2$ termination\cite{Wang2018} while the wafers are etched in order to create a TiO$_2$ termination \cite{Dral2014}. Due to the observation of the 2DEG on as received commercial STO substrate wafers, the 2DEG is generally associated with the TiO$_2$ terminated surface. Under a wide variety of different preparation conditions, the 2DEG is consistently observed with very similar band structure \cite{Hatch2015}.

More generally, perovskite oxides are typically divided into sub-unit cell layers that play different roles, like an active layer that drives the physical phenomena of interest sandwiched between passive doping layers. Cuprate superconductors are a classical example of this type of system and are divided into active superconducting CuO$_2$ layers surrounded by passive charge reservoirs. In a recent experiment, Yan-Feng Lv \textit{et al.} observed dramatically different electronic structure on Bi$_2$Sr$_2$CaCu$_2$O$_{8+\delta}$ (Bi- 2212), a highly studied cuprate superconductor, as they exposed each of the different layers, using Ar+ ion sputtering \cite{Lv2015}. However, the best approach to probe the exotic properties of an active layer is still an open question. Molecular beam epitaxy (MBE) is the ideal tool to allow precise control of different terminations, without the surface damage caused by sputtering.  A shuttered growth approach allows for exploration of a much wider variety of surfaces since we are not restricted to natural crystal cleavage planes. While most research on surface 2DEG of STO assume a TiO$_2$ termination, further research into the pristine SrO termination could lead to new insights into the system.

In this work, we explore the differences in the 2DEG formation between homoepitaxial STO films with SrO and TiO$_2$ terminations by combining synchrotron based ARPES with \textit{in-situ} MBE growth. We observe a clear 2DEG only on SrO terminated STO films and not on films with a TiO$_2$ termination. We explore different growth recipes and find that the accumulation of an extra SrO layer, either from a flux imbalance or a buffer layer at the substrate interface, can cause the STO to continuously rearrange during growth to promote the extra SrO layer to the surface which leads to a 2DEG visible in ARPES.

\section*{Results}
\subsection{MBE Growth \& ARPES Characterization}

We grew homoepitaxial STO using well established methods \cite{Haeni2000} on top of commercially available 0.05\% Nb doped STO (001) substrates. The substrates used in this study were etched and annealed at the vendor to create atomically flat, TiO$_2$ terminated surfaces with well-ordered steps and terraces. We utilized a shuttered approach in which Sr and Ti shutters were opened in turns to impinge on the heated substrate in a background of oxygen. Specific details of the growth parameters can be found in the Methods section. We used reflection high-energy electron diffraction (RHEED) to monitor the quality and thickness of the film during growth. We aligned the RHEED in such a way that the intensity is at a maximum when the Sr shutter closes and a minimum when the Ti shutter closes. A similar alignment methodology is described in detail by H.Y. Sun \textit{et al.} \cite{Sun2018}. We were able to observe large RHEED oscillations, as seen in (Fig. 1a), which are indicative of high quality STO films. \textit{Ex-situ} XPS and XRD measurements, which can be found in the Supplemental Information, further confirm the high quality of the films.

The homoepitaxial, TiO$_2$ terminated STO films are then characterized using ARPES, as shown in (Fig. 1). Each film is exposed to UV light until the spectra intensity is saturated. The background intensity, particularly at higher binding energies increases over time. However the expected 2DEG does not develop near the Fermi Energy, as observed in (Fig 1b). Using our shuttered growth approach, we can terminate the film with an SrO layer by ending with the Sr shutter open last, (Fig. 1d). When SrO terminated films are exposed to UV light, a very bright 2DEG quickly emerges, similar to what has been observed in other measurements on STO, as seen in (Fig 1e). This same phenomenon is observed as well in BTO, where the 2DEG is only visible on the BaO termination, as seen in the Supplemental Fig. 1.

Previous studies indicate unexpected layer ordering and termination when a double SrO layer is deposited during growth \cite{Nie2014, Lee2014a }. To confirm our surface terminations, we perform \textit{ex-situ} XPS on our STO films and the results are summarized in Supplemental Fig. 2. All of the films which show a 2DEG also have similar Sr/Ti ratios. By utilizing the stark contrast in 2DEG formation between the SrO and TiO$_2$ terminations, we can further explore the growth mechanics and termination of various STO recipes. We start by growing a single layer SrO buffer layer, then continue growth using the nominal STO recipe of Sr and Ti deposition in turns. This results in a RHEED oscillation which is 180 degrees \lq out of phase{\rq} compared to a typical STO growth: Sr shutter closes at an intensity minimum and Ti shutter closes at an intensity maximum, (Fig. 1g). ARPES measurements reveal a 2DEG, which along with \textit{ex-situ} XPS, confirms SrO termination despite Ti shutter being open last, as seen in (Fig. 1h).

The correlation between SrO termination and 2DEG formation is robust against disorder. If a double SrO layer is formed mid growth due to off stoichiometric rates, like has been previously reported by Nie, Y.F. \textit{et al.} \cite{Nie2014}, RHEED oscillations flip 180 degrees \lq out of phase{\rq} from nominal STO growth, (Fig. 1j). Provided there are a few good oscillations at the end of growth, the 2DEG is always observed despite the disorder buried a few unit cells beneath the surface, as observed in (Fig. 1k).

\subsection{Partial STO Layers}
Due to our shuttered approach, we can also explore the effect of partial STO layers on the formation of the 2DEG. A succession of partial STO layers were grown on the same film. In between each growth, the sample was photodoped and measured with ARPES with the results shown in Figure 2. All of the partial layer samples reacted when exposed to the UV light. For all of the partial layers, and the fully SrO terminated layer there was intensity which extends up to the Fermi energy. However, only the fully SrO terminated layer had a bright spectra with clear dispersion. These measurements indicate that the formation of the 2DEG is dependent upon a fully SrO terminated layer and not just excess SrO on the surface. When this experiment was repeated with completely separate samples grown for each partial layer termination, it yielded similar results, as seen in the Supplemental Information.

\subsection{Doping Dependence}
We also explored the photon flux dependence of the photodoping response using ARPES and the results can be seen in Figure 3. The SrO terminated homoepitaxial STO films show a weak signal from the very first measurement. This intensity quickly develops into a 2DEG which progressively dopes until a point of saturation (Fig. 3a - e). On the bare substrate, the 2DEG also progressively photodopes, as seen in (Fig. 3f - j). Compared to the fully doped 2DEG on the SrO terminated surface, the intensity of the 2DEG on the bare STO wafer is much weaker. The \textit{in-situ} MBE grown STO films with TiO$_2$ termination show a negligible response over the same range of exposure. However, once we remove our films from vacuum and prepare them using the etching and annealing methods described in Dral, A. P. \textit{et al.} 2014 \cite{Dral2014}, the photodoping response changes. The doping response of the treated TiO$_2$ terminated films is very similar to that of the bare substrate, as seen in (Fig. 3k - o).

\section*{Discussion}

While a 2DEG can form at the surface of a treated STO wafer with TiO$_2$ termination, our results indicate the unexpected importance of the pristine SrO termination. Most research has focused on the TiO$_2$ termination and there are only a few results which explore the role of SrO. There are theories and experiments that suggest for the TiO$_2$ terminated STO, oxygen vacancies cluster near the surface, with the formation barrier being similar for the topmost SrO and TiO$_2$ layers \cite{Sitaputra2015, Jeschke2015}. DFT calculations predict that while the TiO$_2$ terminated surface requires oxygen vacancies to create an accumulation layer for a 2DEG, it should form on a pristine SrO terminated film without the need for photodoping \cite{Sitaputra2015, Delugas2015}. However, the surface morphology of the two terminations are significantly different and disorder can affect the observed energies of in gap states created by oxygen vacancies \cite{Chien2010, Sitaputra2015}. Thus modeling pristine surfaces may not capture the true energy landscape or how inhomogeneity in the different layers affects the oxygen vacancy formation barrier.

Here we propose two scenarios, either of which or both can explain the appearance of the 2DEG only on the pristine SrO termination. Either the SrO termination is the source of the itinerant carriers or its presence causes carriers created elsewhere in the film to accumulate at the surface. Recent experiments suggest the scenario with itinerant charge coming solely from the topmost SrO layer is at least plausible \cite{Walker2015, McKeownWalker2016}. A higher oxygen vacancy formation rate has been observed in fractured STO crystals compared to annealed STO wafers, which suggests exposure of the SrO termination can increase the vacancy formation rate \cite{McKeownWalker2016}. Additionally, modeling x-ray photoelectron spectroscopy data of the mixed Ti$^{3+}$/Ti$^{4+}$ state fits a scenario with all the itinerant charge isolated to the topmost termination layer, but the authors of this study ultimately dismiss this model as nonphysical \cite{Walker2015}.

While we only observe the 2DEG on the SrO terminated surface, both terminations show similar photodoping of the deeper oxygen bands, as seen in the Supplemental Fig. 4. This behavior suggests that some other mechanism is at least partially responsible for the dichotomy of the 2DEG. Recent studies of RHEED patterns during growth have shown a variation of the inner potential between the two surface terminations \cite{Sun2018}. In addition, scanning tunneling spectroscopy studies reveal a 0.25 eV shift in the conduction band onset between the two terminations \cite{Sitaputra2015}. Thus there is evidence for variations in the surface potential despite the traditional assumption that both SrO and TiO$_2$ layers are electrically neutral. In summary, either the topmost SrO layer shifts the surface potential to accumulate electrons at the surface or it is the source of itinerant carriers when oxygen vacancies are created within it. More work is needed to further elucidate the role of the topmost SrO layer in the formation of the observed 2DEG.

The dichotomy observed between the SrO and TiO$_2$ terminations provides interesting insights into lattice structures consisting of active and passive layers. Due to the Ti d orbital character of the observed 2DEG bands, we can assign TiO$_2$ as the active layer in STO, however we do not observe the 2DEG when probing this layer directly. Naively we would expect that directly probing the active layer would provide access to the underlying physics, but our results suggest that the reality is more nuanced. While the SrO layer certainly plays a unique role in STO, it is likely that other systems also require a passive layer termination in order to realize or protect the physics of the active layer.

Our results also suggest that the interface between STO and other materials should be further explored. While the 2DEG on the surface of UV exposed STO is often attributed to the same mechanism as the LAO/STO interface, the later system requires TiO$_2$ at the interface. Understanding the interactions between charge reservoirs and active layers as materials are grown layer by layer could help us to control the relevant emergent properties. Our work helps open new routes to explore the physics of interfaces and achieve a deeper understanding titanates and other oxides materials in general.

\newpage
\begin{figure}
\centerline{
\includegraphics[width= 14cm]{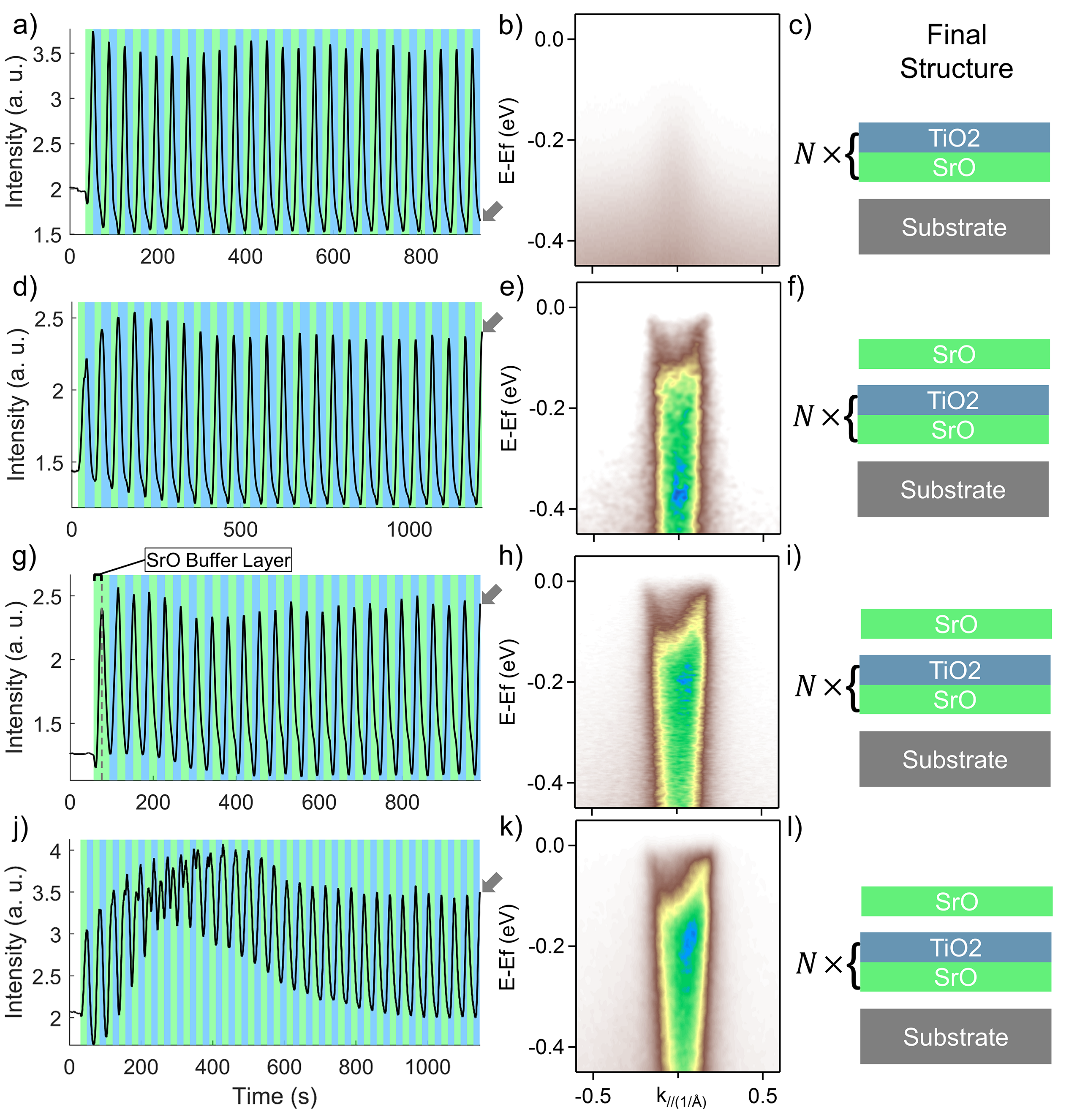}
}
\caption{Growth and ARPES characterization of different STO recipes

Each row of the figure corresponds to different STO growth recipes: Normal TiO$_2$ terminated STO, SrO Capped STO, SrO buffer followed by STO, and Ti deficient STO, respectively. The left column, a), d), g), j) are RHEED oscillations for films with different recipes. The background color represents which shutter was open, blue for Ti and green for Sr. The small arrow marks the intensity at the end of growth. The center column b), e), h), k) are ARPES spectra collected for each recipe once it was fully saturated. Each measurement was taken at a photon energy of 74eV with the sample at 30K.  The right column c), f), i), l) are schematic diagrams which represents the final structure after growth is completed. 
}
\end{figure}

\newpage
\begin{figure}
\centerline{
\includegraphics[width= 14cm]{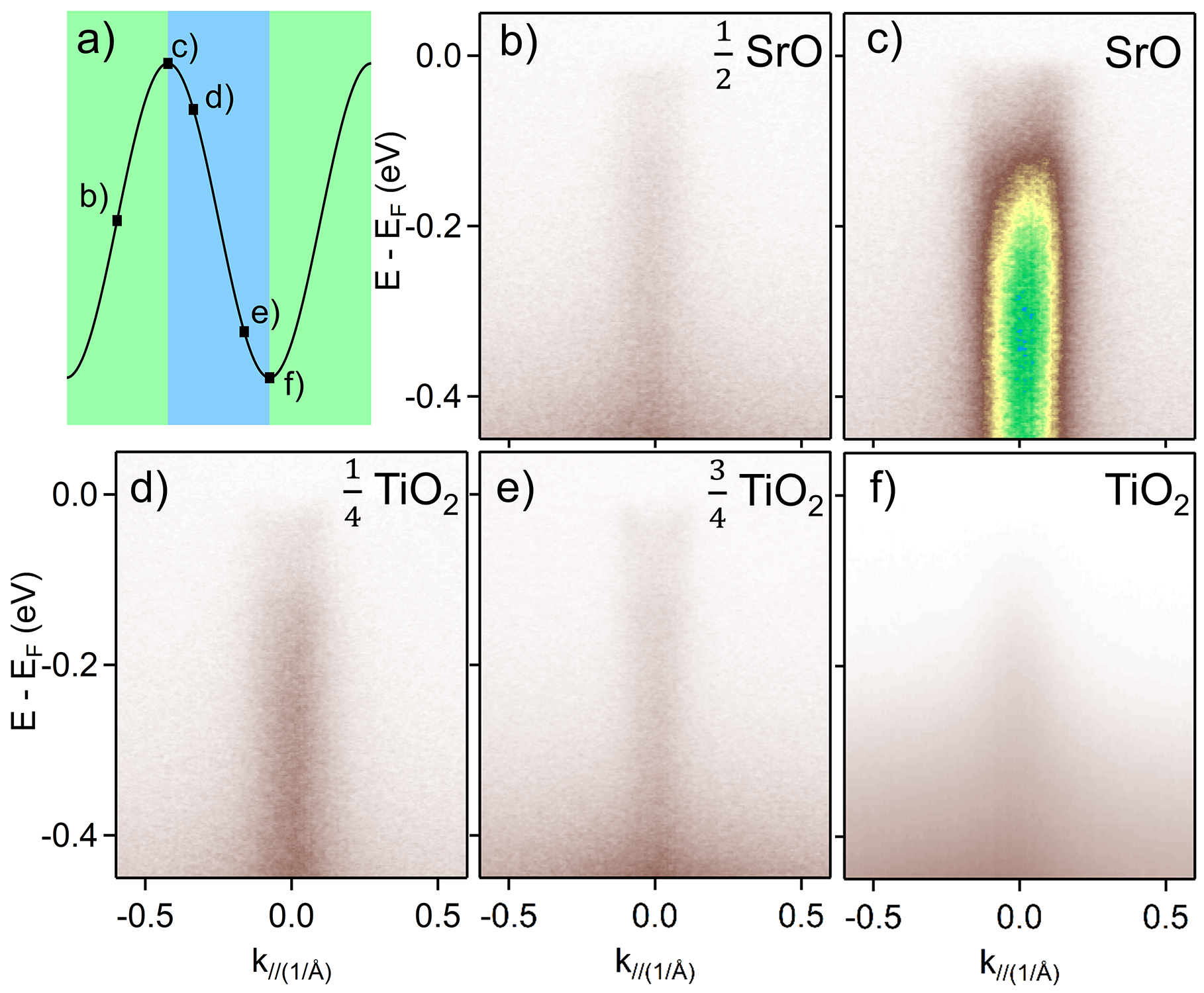}
}
\caption{Partial Layer Growth 

a) Schematic RHEED oscillation to show where along the growth cycle each sample in the Figure was stopped. b) - f) are a series of ARPES spectra taken on the same sample, but stopped at 1/2 SrO layer, full SrO layer, 1/4 TiO$_2$ layer, 3/4 TiO$_2$ layer and full TiO$_2$ layer, respectively, during growth. Each sample was exposed to UV light and fully saturated before collecting the data. The color scale is the same for all spectra. Each measurement was taken at a photon energy of 74eV with the sample at 30K. 
}
\end{figure}

\newpage
\begin{figure}
\centerline{
\includegraphics[width= 14cm]{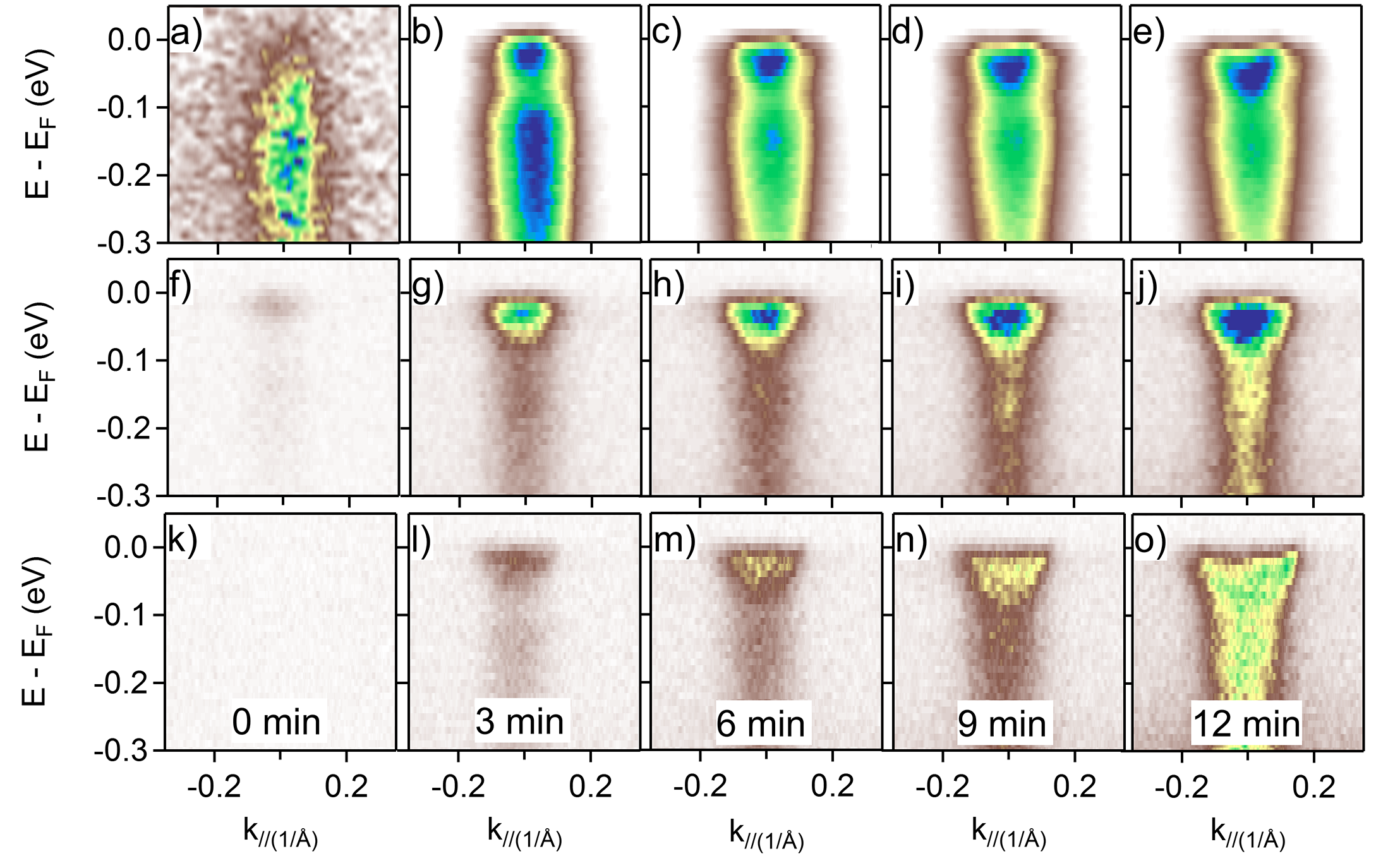}
}
\caption{Photodoping Dependence

Progressive photodoping series for three different samples: SrO termination for a) - e), STO substrate for f) - j), and treated TiO$_2$ terminated MBE grown STO for k) - o). Panels f) - o) have the same color scale. a) - e) are too bright for this same color scale, so the intensity is reduced by a factor of 5. Each spectra was taken using a photon energy of 42eV and with the sample at 30K. Due to the rapid development of the 2DEG with UV exposure, each spectrum was collected in 21 seconds, with ~3 minute spacing between each panel shown. 
}
\end{figure}

\begin{methods}

\subsection{MBE Growth.}
 All of the samples were grown on 0.05\% Nb Doped STEP STO purchased from Shinkosa. The substrates were mounted to inconel sample holders using silver paste. Once loaded into the MBE system, the samples were degassed at $\text{300}^{\circ} \ C$  for 30 minutes. They were then exposed to $6\times 10^{-6}$ torr oxygen partial pressure and heated to $\text{550}^{\circ} \ C$ for growth. The RHEED was aligned along the crystal (110) direction with the electron beam's incident angle increased from zero until the first intensity minimum of the reflected (00) beam was observed. Two cells were used for growth: a differentially pumped source loaded with ultra high purity Sr (99.95\%), and a high temperature cell loaded with ultra high purity Ti (99.995\%). The atomic flux and deposition rates were calibrated and set using a quartz crystal monitor. A shuttered approach was used for growth, which for a typical STO recipe starts with the Sr shutter open first and ends with the Ti shutter open last. The films presented in this paper were grown to 50 u.c. thick, however the same behavior is observed in much thinner films as well. Post growth the samples were cooled down in the oxygen background and then transferred \textit{in-situ} to the ARPES chamber for investigating the electronic structure. 
 \subsection{ARPES}
 All ARPES measurements were conducted at beamline 5-2 of the Stanford Synchrotron Radiation Lightsource. The base pressure of the ARPES chamber was better than $4\times 10^{-11}$ torr. Measurements were taken using linear horizontal polarization, at energies between 32eV and 84eV, while the sample was at 30K. 

\subsection{Ex-situ Characterization}
XPS and XRD were done \textit{ex-situ} at the Stanford Nano Shared Facilities with a Phi Versaprobe and PANalytical X'Pert, respectively.

\end{methods}

\begin{addendum}
\item The authors would like to thank Z. Y. Chen, H. Y. Hwang and B. Moritz for fruitful discussions.  This work was supported by the U.S. Department of Energy, Office of Science, Basic Energy Sciences, Materials Sciences and Engineering Division under contract DE-AC02-76SF00515. Use of the Stanford Synchrotron Radiation Lightsource, SLAC National Accelerator Laboratory, is supported by the U.S. Department of Energy, Office of Science, Office of Basic Energy Sciences under Contract No. DE-AC02-76SF00515. Part of this work was performed at the Stanford Nano Shared Facilities (SNSF), supported by the National Science Foundation under award ECCS-1542152.
\item[Author Contributions] S.N.R., R.G.M., H.S. and T.J. contributed to growth of STO films. S.N.R., T.J. and H.M.S. carried out the ARPES measurements. S.N.R. and T.J. performed the \textit{ex-situ} characterization. M.H. and D.H.L contributed to the operation of ARPES beamline. S.N.R. performed data analysis and wrote the manuscript. Z.X.S. and R.G.M. oversaw the project. All authors reviewed and edited the manuscript.
 
 \item[Competing Interests] The authors declare that they have no
competing financial interests.
 \item[Correspondence] Correspondence and requests for materials
should be addressed to R.G. Moore (email: rgmoore@slac.stanford.edu).
\end{addendum}


\begin{thebibliography}{10}
\expandafter\ifx\csname url\endcsname\relax
  \def\url#1{\texttt{#1}}\fi
\expandafter\ifx\csname urlprefix\endcsname\relax\def\urlprefix{URL }\fi
\providecommand{\bibinfo}[2]{#2}
\providecommand{\eprint}[2][]{\url{#2}}

\bibitem{MullerK.Burkard1979}
\bibinfo{author}{Muller, K.} \& \bibinfo{author}{Burkard, H.}
\newblock \bibinfo{title}{{SrTiO$_3$: An intrinsic quantum paraelectric below 4
  K}}.
\newblock \emph{\bibinfo{journal}{Physical Review B}}
  \textbf{\bibinfo{volume}{19}} (\bibinfo{year}{1979}).


\bibitem{Temperatures1967}
\bibinfo{author}{Koonce, C.~S.} \& \bibinfo{author}{Cohen, M.~L.}
\newblock \bibinfo{title}{{Superconducting Transition Temperatures of
  Semiconducting SrTiO3}}.
\newblock \emph{\bibinfo{journal}{Physical Review}}
  \textbf{\bibinfo{volume}{163}} (\bibinfo{year}{1967}).

\bibitem{Wang2012}
\bibinfo{author}{Wang, Q.~Y.} \emph{et~al.}
\newblock \bibinfo{title}{{Interface-induced high-temperature superconductivity
  in single unit-cell FeSe films on SrTiO$_3$}}.
\newblock \emph{\bibinfo{journal}{Chinese Physics Letters}}
  \textbf{\bibinfo{volume}{29}} (\bibinfo{year}{2012}).


\bibitem{Lee2014}
\bibinfo{author}{Lee, J.~J.} \emph{et~al.}
\newblock \bibinfo{title}{{Interfacial mode coupling as the origin of the
  enhancement of T$_c$ in FeSe films on SrTiO$_3$}}.
\newblock \emph{\bibinfo{journal}{Nature}} \textbf{\bibinfo{volume}{515}},
  \bibinfo{pages}{245--248} (\bibinfo{year}{2014}).


\bibitem{Ohtomo2004}
\bibinfo{author}{Ohtomo, A.} \& \bibinfo{author}{Hwang, H.~Y.}
\newblock \bibinfo{title}{{A high-mobility electron gas at the LaAlO$_3$ / SrTiO$_3$
  heterointerface}}.
\newblock \emph{\bibinfo{journal}{Nature}} \textbf{\bibinfo{volume}{427}},
  \bibinfo{pages}{423--427} (\bibinfo{year}{2004}).

\bibitem{Caviglia2008}
\bibinfo{author}{Caviglia, A.~D.} \emph{et~al.}
\newblock \bibinfo{title}{{Electric Field Control of the LaAlO$_3$
  /SrTiO$_3$ Interface Ground State}}.
\newblock \emph{\bibinfo{journal}{Nature}} \textbf{\bibinfo{volume}{456}},
  \bibinfo{pages}{2--5} (\bibinfo{year}{2008}).


\bibitem{Bert2011}
\bibinfo{author}{Bert, J.~A.} \emph{et~al.}
\newblock \bibinfo{title}{{Direct imaging of the coexistence of ferromagnetism
  and superconductivity at the LaAlO$_3$/SrTiO$_3$ interface}}.
\newblock \emph{\bibinfo{journal}{Nature Physics}}
  \textbf{\bibinfo{volume}{7}}, \bibinfo{pages}{767--771}
  (\bibinfo{year}{2011}).


\bibitem{Li2011}
\bibinfo{author}{Li, L.}, \bibinfo{author}{Richter, C.},
  \bibinfo{author}{Mannhart, J.} \& \bibinfo{author}{Ashoori, R.~C.}
\newblock \bibinfo{title}{{Coexistence of magnetic order and two-dimensional
  superconductivity at LaAlO$_3$/SrTiO$_3$ interfaces}}.
\newblock \emph{\bibinfo{journal}{Nature Physics}}
  \textbf{\bibinfo{volume}{7}}, \bibinfo{pages}{762--766}
  (\bibinfo{year}{2011}).


\bibitem{Meevasana2010}
\bibinfo{author}{Meevasana, W.} \emph{et~al.}
\newblock \bibinfo{title}{{Strong energy-momentum dispersion of phonon-dressed
  carriers in the lightly doped band insulator SrTiO$_3$}}.
\newblock \emph{\bibinfo{journal}{New Journal of Physics}}
  \textbf{\bibinfo{volume}{12}}, \bibinfo{pages}{0--11} (\bibinfo{year}{2010}).


\bibitem{Santander-Syro2011}
\bibinfo{author}{Santander-Syro, A.~F.} \emph{et~al.}
\newblock \bibinfo{title}{{Two-dimensional electron gas with universal subbands
  at the surface of SrTiO$_3$}}.
\newblock \emph{\bibinfo{journal}{Nature}} \textbf{\bibinfo{volume}{469}},
  \bibinfo{pages}{189--194} (\bibinfo{year}{2011}).


\bibitem{Hatch2015}
\bibinfo{author}{Hatch, R.~C.}, \bibinfo{author}{Choi, M.},
  \bibinfo{author}{Posadas, A.~B.} \& \bibinfo{author}{Demkov, A.~A.}
\newblock \bibinfo{title}{{Comparison of acid- and non-acid-based surface
  preparations of Nb-doped SrTiO$_3$(001)}}.
\newblock \emph{\bibinfo{journal}{Journal of Vacuum Science {\&} Technology B,
  Nanotechnology and Microelectronics: Materials, Processing, Measurement, and
  Phenomena}} \textbf{\bibinfo{volume}{33}}, \bibinfo{pages}{061204}
  (\bibinfo{year}{2015}).


\bibitem{Walker2015}
\bibinfo{author}{Walker, S. M.~K.} \emph{et~al.}
\newblock \bibinfo{title}{{Carrier-Density Control of the SrTiO$_3$(001) Surface
  2D Electron Gas studied by ARPES}}.
\newblock \emph{\bibinfo{journal}{Advanced Materials}}
  \textbf{\bibinfo{volume}{27}}, \bibinfo{pages}{3894--3899}
  (\bibinfo{year}{2015}).
  

\bibitem{Sitaputra2015}
\bibinfo{author}{Sitaputra, W.}, \bibinfo{author}{Sivadas, N.},
  \bibinfo{author}{Skowronski, M.}, \bibinfo{author}{Xiao, D.} \&
  \bibinfo{author}{Feenstra, R.~M.}
\newblock \bibinfo{title}{Oxygen vacancies on SrO-terminated SrTiO$_3$(001)
  surfaces studied by scanning tunneling spectroscopy}.
\newblock \emph{\bibinfo{journal}{Physical Review B}}
  \textbf{\bibinfo{volume}{91}}, \bibinfo{pages}{205408}
  (\bibinfo{year}{2015}).

\bibitem{Wang2014}
\bibinfo{author}{Wang, Z.} \emph{et~al.}
\newblock \bibinfo{title}{{Anisotropic two-dimensional electron gas at
  SrTiO$_3$(110)}}.
\newblock \emph{\bibinfo{journal}{Proceedings of the National Academy of
  Sciences}} \textbf{\bibinfo{volume}{111}}, \bibinfo{pages}{3933--3937}
  (\bibinfo{year}{2014}).
  

\bibitem{McKeownWalker2016}
\bibinfo{author}{{McKeown Walker}, S.} \emph{et~al.}
\newblock \bibinfo{title}{{Absence of giant spin splitting in the
  two-dimensional electron liquid at the surface of SrTiO$_3$ (001)}}.
\newblock \emph{\bibinfo{journal}{Physical Review B}}
  \textbf{\bibinfo{volume}{93}}, \bibinfo{pages}{1--5} (\bibinfo{year}{2016}).


\bibitem{Zhang2017}
\bibinfo{author}{Zhang, C.} \emph{et~al.}
\newblock \bibinfo{title}{{Ubiquitous strong electron-phonon coupling at the
  interface of FeSe/SrTiO$_3$}}.
\newblock \emph{\bibinfo{journal}{Nature Communications}}
  \textbf{\bibinfo{volume}{8}}, \bibinfo{pages}{1--6} (\bibinfo{year}{2017}).


\bibitem{McKeownWalker2014}
\bibinfo{author}{{McKeown Walker}, S.} \emph{et~al.}
\newblock \bibinfo{title}{{Control of a Two-Dimensional Electron Gas on SrTiO$_3$
  (111) by Atomic Oxygen}}.
\newblock \emph{\bibinfo{journal}{Physical Review Letters}}
  \textbf{\bibinfo{volume}{113}}, \bibinfo{pages}{1--5} (\bibinfo{year}{2014}).


\bibitem{Wang2018}
\bibinfo{author}{Wang, A.} \& \bibinfo{author}{Chien, T.~Y.}
\newblock \bibinfo{title}{{Perspectives of cross-sectional scanning tunneling
  microscopy and spectroscopy for complex oxide physics}}.
\newblock \emph{\bibinfo{journal}{Physics Letters, Section A: General, Atomic
  and Solid State Physics}} \textbf{\bibinfo{volume}{382}},
  \bibinfo{pages}{739--748} (\bibinfo{year}{2018}).


\bibitem{Dral2014}
\bibinfo{author}{Dral, A.~P.} \emph{et~al.}
\newblock \bibinfo{title}{{Atomically Defined Templates for Epitaxial Growth of
  Complex Oxide Thin Films}}.
\newblock \emph{\bibinfo{journal}{Journal of Visualized Experiments}}
  \bibinfo{pages}{1--13} (\bibinfo{year}{2014}).


\bibitem{Lv2015}
\bibinfo{author}{Lv, Y.~F.} \emph{et~al.}
\newblock \bibinfo{title}{{Mapping the Electronic Structure of Each Ingredient
  Oxide Layer of High- T$_c$ Cuprate Superconductor Bi$_2$Sr$_2$CaCu$_2$O$_{8+\delta}$}}.
\newblock \emph{\bibinfo{journal}{Physical Review Letters}}
  \textbf{\bibinfo{volume}{115}}, \bibinfo{pages}{1--5} (\bibinfo{year}{2015}).


\bibitem{Haeni2000}
\bibinfo{author}{Haeni, J.~H.}, \bibinfo{author}{Theis, C.~D.} \&
  \bibinfo{author}{Schlom, D.~G.}
\newblock \bibinfo{title}{{RHEED Intensity Oscillations for the Stoichiometric
  Growth of SrTiO$_3$ Thin Films by Reactive Molecular Beam Epitaxy}}.
\newblock \emph{\bibinfo{journal}{Journal of Electroceramics}}
  \textbf{\bibinfo{volume}{4}}, \bibinfo{pages}{385--391}
  (\bibinfo{year}{2000}).


\bibitem{Sun2018}
\bibinfo{author}{Sun, H.~Y.} \emph{et~al.}
\newblock \bibinfo{title}{{Chemically specific termination control of oxide
  interfaces via layer-by-layer mean inner potential engineering}}.
\newblock \emph{\bibinfo{journal}{Nature Communications}}
  \textbf{\bibinfo{volume}{9}} (\bibinfo{year}{2018}).


\bibitem{Nie2014}
\bibinfo{author}{Nie, Y.~F.} \emph{et~al.}
\newblock \bibinfo{title}{{Atomically precise interfaces from
  non-stoichiometric deposition}}.
\newblock \emph{\bibinfo{journal}{Nature Communications}}
  \textbf{\bibinfo{volume}{5}} (\bibinfo{year}{2014}).


\bibitem{Lee2014a}
\bibinfo{author}{Lee, J.~H.} \emph{et~al.}
\newblock \bibinfo{title}{{Dynamic layer rearrangement during growth of layered
  oxide films by molecular beam epitaxy}}.
\newblock \emph{\bibinfo{journal}{Nature Materials}}
  \textbf{\bibinfo{volume}{13}}, \bibinfo{pages}{879--883}
  (\bibinfo{year}{2014}).


\bibitem{Jeschke2015}
\bibinfo{author}{Jeschke, H.~O.}, \bibinfo{author}{Shen, J.} \&
  \bibinfo{author}{VAlenti, R.}
\newblock \bibinfo{title}{{Localized versus itinerant states created by
  multiuple oxygen vacancies in SrTiO$_3$}}.
\newblock \emph{\bibinfo{journal}{New Journal of Physics}}
  \textbf{\bibinfo{volume}{17}}, \bibinfo{pages}{023034}
  (\bibinfo{year}{2015}).


\bibitem{Delugas2015}
\bibinfo{author}{Delugas, P.}, \bibinfo{author}{Fiorentini, V.},
  \bibinfo{author}{Mattoni, A.} \& \bibinfo{author}{Filippetti, A.}
\newblock \bibinfo{title}{{Intrinsic origin of two-dimensional electron gas at
  the (001) surface of SrTiO$_3$}}.
\newblock \emph{\bibinfo{journal}{Physical Review B - Condensed Matter and
  Materials Physics}} \textbf{\bibinfo{volume}{91}}, \bibinfo{pages}{1--12}
  (\bibinfo{year}{2015}).


\bibitem{Chien2010}
\bibinfo{author}{Chien, T.}, \bibinfo{author}{Guisinger, N.~P.} \&
  \bibinfo{author}{Freeland, J.~W.}
\newblock \bibinfo{title}{{Survey of fractured SrTiO$_3$ surfaces: From the
  micrometer to nanometer scale}}.
\newblock \emph{\bibinfo{journal}{Journal of Vacuum Science {\&} Technology B}}
  \textbf{\bibinfo{volume}{28}}, \bibinfo{pages}{C5A11} (\bibinfo{year}{2010}).

\end{thebibliography}
\end{document}